\documentclass[english,aps,pre,preprint]{revtex4}
\usepackage[latin9]{inputenc}
\usepackage{babel}
\usepackage{bm}
\usepackage{graphicx}
\usepackage[unicode=true,pdfusetitle,
 bookmarks=true,bookmarksnumbered=false,bookmarksopen=false,
 breaklinks=false,pdfborder={0 0 1},backref=false,colorlinks=false]
 {hyperref}
 
\usepackage{tikz}

\makeatletter
\@ifundefined{textcolor}{}
{%
 \definecolor{BLACK}{gray}{0}
 \definecolor{WHITE}{gray}{1}
 \definecolor{RED}{rgb}{1,0,0}
 \definecolor{GREEN}{rgb}{0,1,0}
 \definecolor{BLUE}{rgb}{0,0,1}
 \definecolor{CYAN}{cmyk}{1,0,0,0}
 \definecolor{MAGENTA}{cmyk}{0,1,0,0}
 \definecolor{YELLOW}{cmyk}{0,0,1,0}
}

\makeatother

\begin{document}

\title{Kinetic diffuse boundary condition for high-order lattice
Boltzmann model with streaming-collision mechanism }

\author{Jianping meng}

\email{jianping.meng@strath.ac.uk}

\affiliation{Department of Mechanical \& Aerospace Engineering, University of
Strathclyde, Glasgow G1 1XJ, UK}

\author{Yonghao Zhang}

\email{yonghao.zhang@strath.ac.uk}

\affiliation{Department of Mechanical \& Aerospace Engineering, University of
Strathclyde, Glasgow G1 1XJ, UK}
\begin{abstract}
The implementation of the kinetic diffuse boundary condition with the characteristic streaming-collision mechanism is studied for the high-order lattice Boltzmann (LB) models. The obtained formulation is also tested and validated numerically
for three high-order LB models for both isothermal and thermal Couette
flows. The  streaming-collision mechanism ensures that high-order LB models can
retain particle feature while go beyond the Navier-Stokes hydrodynamics. 
\end{abstract}

\maketitle

\section{Introduction}

High-order models have recently attracted considerable interests in
the lattice Boltzmann (LB) community. For these models, high-order
terms in the expanded distribution function are retained, thus multi-speed
lattices have to be used. By doing so, we have a few benefits,
e.g. consistent description of thermal flows, the Galilean invariance
of the transport coefficients, improved model capability for compressible
and rarefied flows\cite{2006JFM550413S,Meng2011a,0295-5075-81-3-34005}.
Meanwhile, high-order models can still preserve the simplicity of the standard LB model. High-order LB models are tererfore often applied to thermal flows, compressible
flows, and rarefied flows \cite{Nie_Shan_Chen_2009,Meng2012,2006JFM550413S}.

Due to their kinetic origin, high-order LB models
have shown to be able to approach the Boltzmann model equation such as the Bhatnagar-Gross-Krook (BGK) equation by increasing
the expansion and quadrature order\cite{2006JFM550413S}. Numerically,
it is found that high-order models are capable of describing rarefaction
effects up to the early transitional regime with slightly increased
discrete velocities. For example, the velocity-slip can be captured by
a D2Q16 model for the Knudsen number up to $0.5$ where D$m$Q$n$ refers to the standard $m$ dimenional and $n$ discrete velocities. A recent study
on thermal flows shows that thermal rarefaction effects can also
be captured by using a moderate number of discrete velocities with an affordable
computational cost\cite{Meng2012}. In addition to capturing the non-equilibrium
effects, high-order models are also useful for other applications.
For instance, by using a D2Q25 model, Daniel et at\cite{Lycett-Brown2011}
have shown that high Weber number can be achievable for droplet collision
simulations. 

Implementation of boundary conditions is crucial to application
of high-order LB models. The kinetic diffuse-reflection boundary condition
has shown to be able to predict velocity-slip and temperature-jump
at the solid boundary.
Moreover, the positivity of distribution function can always be maintained
(provided that the distribution function from the bulk is positive),
which is key to numerical stability. However, due to the multi-speed
lattice of high-order models, the formulation of the kinetic diffuse-reflection
boundary condition with the characteristic ``streaming and collision'' mechanism
is yet to be developed. The successful implementations so far for
high-order models are based on various finite difference scheme \cite{Sofonea2009,PhysRevE.79.066706},
where the highly desirable ``streaming and collision'' mechanism
disappears. Because of this ``streaming and collision'' mechanism, LB method is often regarded a particle method. The purpose of this work is to formulate the kinetic diffuse reflection boundary condition to retain the ``streaming and collision'' feature.

\section{Brief description of high order lattice Boltzmann model}

The Boltzmann-BGK equation can be dicretized using a systematic procedure (see
\cite{2006JFM550413S,PhysRevLett.80.65,He1997} for the details) to derive
the LB governing equation, which can be written as 

\begin{equation}
\frac{\partial f_{\alpha}}{\partial t}+c_{\alpha,i}\frac{\partial f_{\alpha}}{\partial x_{i}}=-\frac{1}{\tau}\left(f_{\alpha}-f_{\alpha}^{eq}\right),\label{lbgk}
\end{equation}
where $f$ denotes the single-particle distribution function evaluated
at a discrete velocity $\bm{c}_{\alpha}$, $f^{eq}$ is the truncated
Maxwellian distribution, while $\tau$ is the mean relaxation time.
For convenience, a non-dimensional system 

\begin{equation}
x_{k}=\frac{\hat{x}_{k}}{L},u_{k}=\frac{\hat{u}_{k}}{\sqrt{RT_{0}}},t=\frac{\sqrt{RT_{0}}\hat{t}}{L},c_{k}=\frac{\hat{c}{}_{k}}{\sqrt{RT_{0}}},T=\frac{\hat{T}}{T_{0}},\tau=\frac{\sqrt{RT_{0}}\hat{\tau}}{L}
\end{equation}
\[
f=\frac{\hat{f}(RT_{0})^{D/2}}{\rho_{0}},\rho=\frac{\hat{\rho}}{\rho_{0}},p=\frac{\hat{p}}{p_{0}},\mu=\frac{\hat{\mu}}{\mu_{0}},q_{i}=\frac{\hat{q}_{i}}{p_{0}\sqrt{RT_{0}}},\sigma_{ij}=\frac{\sigma_{ij}}{p_{0}}.
\]
can be introduced in Eq.(\ref{lbgk}), where the symbols with hat represent
the dimensional quantities. The common notations are used to represent
physical quantities, i.e. $\rho$ denotes density; $\mu$, dynamic viscosity; $u$, velocity; $p$, pressure; $T$, temperature; $\sigma$, stress and
$q$, heat flux. $L$ is the characteristic length of the system.
The symbols with subscript $0$ are the corresponding reference quantities.
With this non-dimensional system, the equation of state becomes $p=\rho T$.
The mean relaxation time can be written explicitly as $\tau=\mu_{0}\sqrt{RT_{0}}\mu/(p_{0}Lp)$,
which is related to the viscosity and pressure. Meanwhile, the macroscopic
quantities can be obtained as

\begin{equation}
\left[\begin{array}{c}
\rho\\
\rho u_{i}\\
\sigma_{ij}\\
q_{i}\\
\rho DT
\end{array}\right]=\sum_{\alpha=1}^{d}f_{\alpha}\left[\begin{array}{c}
1\\
c_{\mbox{\ensuremath{\alpha},}i}\\
C_{<\mbox{\ensuremath{\alpha},}i}C_{\mbox{\ensuremath{\alpha},}j>}\\
\frac{1}{2}C_{\mbox{\ensuremath{\alpha},}i}C_{\mbox{\ensuremath{\alpha},}i}C_{\mbox{\ensuremath{\alpha},}j}\\
C_{\mbox{\ensuremath{\alpha},}i}C_{\mbox{\ensuremath{\alpha},}i}
\end{array}\right],\label{eq:summac}
\end{equation}
where $D$ is the space dimension number and $d$ is the total discrete
velocity number and the angle brackets $<\cdots>$ indicates the trace-free
part of the tensor. The discrete velocity $\bm{c}_{\alpha}$ and its
weights $w_{\alpha}$ may be determined through several ways, e.g.,
\cite{2006JFM550413S,PhysRevE.79.046701,PhysRevE.81.036702} list explicitly
various orders of discrete velocity sets. For
convenience, we use the notation $\bm{\xi}=\{\bm{c}_{\alpha},w_{\alpha}\},\alpha=1..d$
to represent the discrete velocity set. With an appropriate $\bm{\xi}$,
the explicit form of the truncated Maxwellian distribution can be
given as
\begin{eqnarray}
f_{\alpha}^{eq} & = & \rho g_{\alpha}^{eq}\nonumber \\
g_{\alpha}^{eq} & = & w_{\alpha}\left\{ 1+c_{i}u_{i}+\frac{1}{2}\left[(c_{i}u_{i})^{2}-u_{i}u_{i}+(T-1)(c_{i}c_{i}-D)\right]\right.\label{eq:feq}\\
 & + & \frac{c_{i}u_{i}}{6}[(c_{i}u_{i})^{2}-3u_{i}u_{i}+3(T-1)(c_{i}c_{i}-D-2)]\nonumber \\
 & + & \frac{1}{24}[(c_{i}u_{i})^{4}-6(u_{i}c_{i})^{2}u_{j}u_{j}+3(u_{j}u_{j})^{2}]\nonumber \\
 & + & \frac{T-1}{4}[(c_{i}c_{i}-D-2)((u_{i}c_{i})^{2}-u_{i}u_{i})-2(u_{i}c_{i})^{2}]\nonumber \\
 & + & \frac{(T-1)^{2}}{8}\left.\left[(c_{i}c_{i})^{2}-2(D+2)c_{i}c_{i}+D(D+2)\right]\right\} ,\nonumber 
\end{eqnarray}
in which the forth order expansion is used. If an isothermal flow
is concerned, the temperature $T$ should be set to $1$. Moreover,
to simulate incompressible flows, it is common to use only the second
order terms of $g_{\alpha}^{eq}$. 

With Eq.(\ref{lbgk}) and (\ref{eq:feq}), the final issue is to choose
an appropriate numerical scheme. Following the spirit of ``streaming
and collision'' mechanism \cite{He1998282}, an implicit scheme 

\begin{eqnarray*}
f_{\alpha}(\bm{x}+\bm{c}_{\alpha}dt,t+dt)-f(\bm{x},t) & = & \frac{dt}{2\tau(\bm{x},t)}\left[f_{\alpha}^{eq}(\bm{x},t)-f_{\alpha}(\bm{x},t)\right]\\
 & + & \frac{dt}{2\tau(\bm{x}+\bm{c}_{\alpha}dt,t+dt)}\left[f_{\alpha}^{eq}(\bm{x}+\bm{c}_{\alpha}dt,t+dt)-f_{\alpha}(\bm{x}+\bm{c}_{\alpha}dt,t+dt)\right],
\end{eqnarray*}
can be constructed. If we introduce a new variable

\[
\tilde{f}_{\alpha}=f_{\alpha}+\frac{dt}{2\tau}(f_{\alpha}-f_{\alpha}^{eq})
\]
to eliminate the implicitness, we will get the evolution equation
for$\tilde{f}$ as

\begin{equation}
\tilde{f}_{\alpha}(\bm{x}+\bm{c}_{\alpha}dt,t+dt)-\tilde{f}_{\alpha}(\bm{x},t)=-\frac{dt}{\tau(\bm{x},t)+0.5dt}\left[\tilde{f}_{\alpha}(\bm{x},t)-f_{\alpha}^{eq}(\bm{\bm{x}},t)\right].\label{eq:scheme}
\end{equation}
The advantage of Eq.(\ref{eq:scheme}) is that, if the discrete velocities
are tied to discretization of the space and time by choosing $\bm{\xi}$
with integer value, the evolution of $\tilde{f_{\alpha}}$ can be
accomplished in a way similar to a ``particle'', which makes the
LB method simple but still flexible.

With the variable $\tilde{f_{\alpha}}$, conservative quantities like
density can still be obtained by using Eq.(\ref{eq:summac}) without
changing form but some conversions are needed for shear stress and
heat flux (cf. \cite{He1998282}). In addition, the mean relaxation
time $\tau$ may be related to the local gas temperature for thermal
problems.

\section{Kinetic diffuse-reflection-type boundary condition}

The essential idea of the kinetic diffuse-reflection boundary condition
is that an outgoing particle completely forgets its history and its
velocity is re-normalized by the Maxwellian distribution. To implement
this diffuse-reflection principle for the high-order LB model, we
will follow the procedure described in \cite{Cercignani2000,PhysRevE.66.026311,gatignol:2022}.
Moreover, the discussion is based on the assumption that the effective
particle-wall interaction time is small compared to any characteristic
time of interest and no permanent adsorption occurs\cite{Cercignani2000}. 

For the high-order models, as ``particles'' from more than one layer
of computational grids can hit the wall, we have to properly identify
them in order to implement the boundary condition. For this purpose,
$N$ layers of ghost grids are introduced (see the example of the
D2Q17 lattice and its grid arrangement shown in Fig.\ref{fig:D2Q17-Lattice}
and \ref{fig:Illustration}), where $N$ can be determined via the
corresponding maximum value of the discrete velocity heading towards
the wall (e.g., $N=3$ for the D2Q17 lattice). As a common practice,
the physical wall is located at the half grid space between the ghost
and fluid grids. To further distinguish incoming and outgoing particles,
we use $\bm{c}_{\alpha,l}^{\prime}$ and $\bm{c}_{\alpha,l}$ to represent
their velocities respectively, where $l$ denotes the layer number
of the ghost grid ranging from $0$ to $N-1$. Similarly, the distributions
of incoming and outgoing particles at layer $l$ are written as $f_{\alpha,l}^{I}(\bm{x}_{w},t)$$ $
and $f_{\alpha,l}^{O}(\bm{x}_{w},t)$, where the superscripts $I$
and $O$ stand for `incoming' and `outgoing'. The corresponding
discrete velocities must satisfy the condition $(\bm{c}_{\alpha,l}^{\prime}-\bm{u}_{w})\cdot\bm{n}dt<-l dx$
and $(\bm{c}_{\alpha,l}-\bm{u}_{w})\cdot\bm{n}dt>ldx$, $\bm{n}$ denotes
the unit vector normal to the wall surface $\partial\Omega$ at $\bm{x}$
and directed from the wall into the gas. Note in the present lattice system $dx/dt=1$ so the conditions are equivalent to  $(\bm{c}_{\alpha,l}^{\prime}-\bm{u}_{w})\cdot\bm{n}<-l$
and $(\bm{c}_{\alpha,l}-\bm{u}_{w})\cdot\bm{n}>l$.   Indeed, $\bm{c}_{\alpha,l}^{\prime}$
and $\bm{c}_{\alpha,l}$ are a symmetric pair. On the other hand,
the known information of the wall, i.e., the position, velocity and
temperature, are represented by $\bm{x}_{w}$,$\bm{u}_{w}$ and $T_{w}$. 

Obviously, the distribution $f_{\alpha,l}^{I}(\bm{x}_{w},t)$$ $
can be obtained by naturally streaming the distribution function at
fluid grids into the corresponding ghost ones. We need to determine
the unknown distribution $f_{\alpha,l}^{O}(\bm{x}_{w},t)$ according
to the principle of diffusion reflection. Similar to the derivation
of the continuum version of diffusion-reflection condition \cite{Cercignani2000},
we first write down the mass of outgoing and incoming particles as, 

\begin{equation}
\mathcal{M}_{\alpha,l}^{O}=f_{\alpha,l}^{O}(\bm{x}_{w},t)dV\bm{\quad x}\in\partial\Omega,(\bm{c}_{\alpha,l}-\bm{u}_{w})\cdot\bm{n}>l,
\end{equation}

\begin{equation}
\mathcal{M}_{\alpha,l}^{I}=f_{\alpha,l}^{I}(\bm{x}_{w},t)dV\quad\bm{x}\in\partial\Omega,(\bm{c}_{\alpha,l}^{\prime}-\bm{u}_{w})\cdot\bm{n}<-l,
\end{equation}
where $\mathcal{M}$ stands for mass and $dV$ denotes the volume
of the grid cell. It is worth noting again here that, due to the exact
advection of the LB method (cf. Eq.(\ref{eq:scheme})), the flux term
in the continuum version (cf. Eq. (1.11.1) in \cite{Cercignani2000}),
can be replaced by the distribution function itself. Hence, according to
the mass conservation, we have,

\begin{equation}
\mathcal{M}_{\alpha,l}^{O}=\sum_{l}^{N-1}\,\sum_{(\bm{c}_{\alpha,l}^{\prime}-\bm{u}_{w})\cdot\bm{n}<-l}R(\bm{c}_{\alpha,l}^{\prime}\rightarrow\bm{c}_{\alpha,l},\bm{x}_{w},t)\mathcal{M}_{\alpha,l}^{I},
\end{equation}
where $R(\bm{c}_{\alpha,l}^{\prime}\rightarrow\bm{c}_{\alpha,l},\bm{x}_{w},t)$
is the so-called scattering probability. Immediately, we arrive at 

\begin{equation}
f_{\alpha,l}^{O}=\sum_{l=0}^{N-1}\,\sum_{(\bm{c}_{\alpha,l}^{\prime}-\bm{u}_{w})\cdot\bm{n}<-l}R(\bm{c}_{\alpha,l}^{\prime}\rightarrow\bm{c}_{\alpha,l},\bm{x}_{w},t)f_{\alpha,l}^{I}(\bm{x}_{w},t).
\end{equation}
Moreover, the scattering probability $R$ must satisfy the property
of non-negativeness, normalization and reciprocity condition\cite{Cercignani2000}.
Particularly, the normalization condition, corresponding to mass conservation
under the assumption of no permanent adsorption, can be written as,

\begin{equation}
\sum_{l=0}^{N-1}\,\sum_{(\bm{c}_{\alpha,l}-\bm{u}_{w})\cdot\bm{n}>l}R(\bm{c}_{\alpha,l}^{\prime}\rightarrow\bm{c}_{\alpha,l},\bm{x}_{w},t)=1.
\end{equation}
So far, the discussion is still generic as we have not introduced
any specific assumption for the diffuse-reflection principle. Therefore,
the above formulation may also be used to derive other type of boundary
condition. 

If the assumption of the diffuse-reflection boundary condition is
applied, the scattering probability can be easily calculated as 

\begin{equation}
R_{\mathcal{D}}\bm{(c}_{\alpha,l}^{\prime}\rightarrow\bm{c}_{\alpha,l})=\frac{g_{\alpha}^{eq}(\bm{u}_{w},T_{w})}{\sum_{l=0}^{N-1}\,\sum_{(\bm{c}_{\alpha,l}-\bm{u}_{w})\cdot\bm{n}>l}g_{\alpha}^{eq}(\bm{u}{}_{w},T_{w})}.
\end{equation}
Hence, the distributions of outgoing particles can be written as 
\begin{equation}
f_{\alpha,l}^{O}=\frac{\sum_{l=0}^{N-1}\,\sum_{(\bm{c}_{\alpha,l}^{\prime}-\bm{u}_{w})\cdot\bm{n}<-l}f_{\alpha,l}^{I}}{\sum_{l=0}^{N-1}\,\sum_{(\bm{c}_{\alpha,l}-\bm{u}_{w})\cdot\bm{n}>l}g_{\alpha}^{eq}(\bm{u}{}_{w},T_{w})}g_{\alpha}^{eq}(\bm{u}_{w},T_{w}).\label{eq:dkboundary}
\end{equation}

\section{Numerical validation}

\begin{figure}
\begin{centering}
\includegraphics[width=0.3\textwidth]{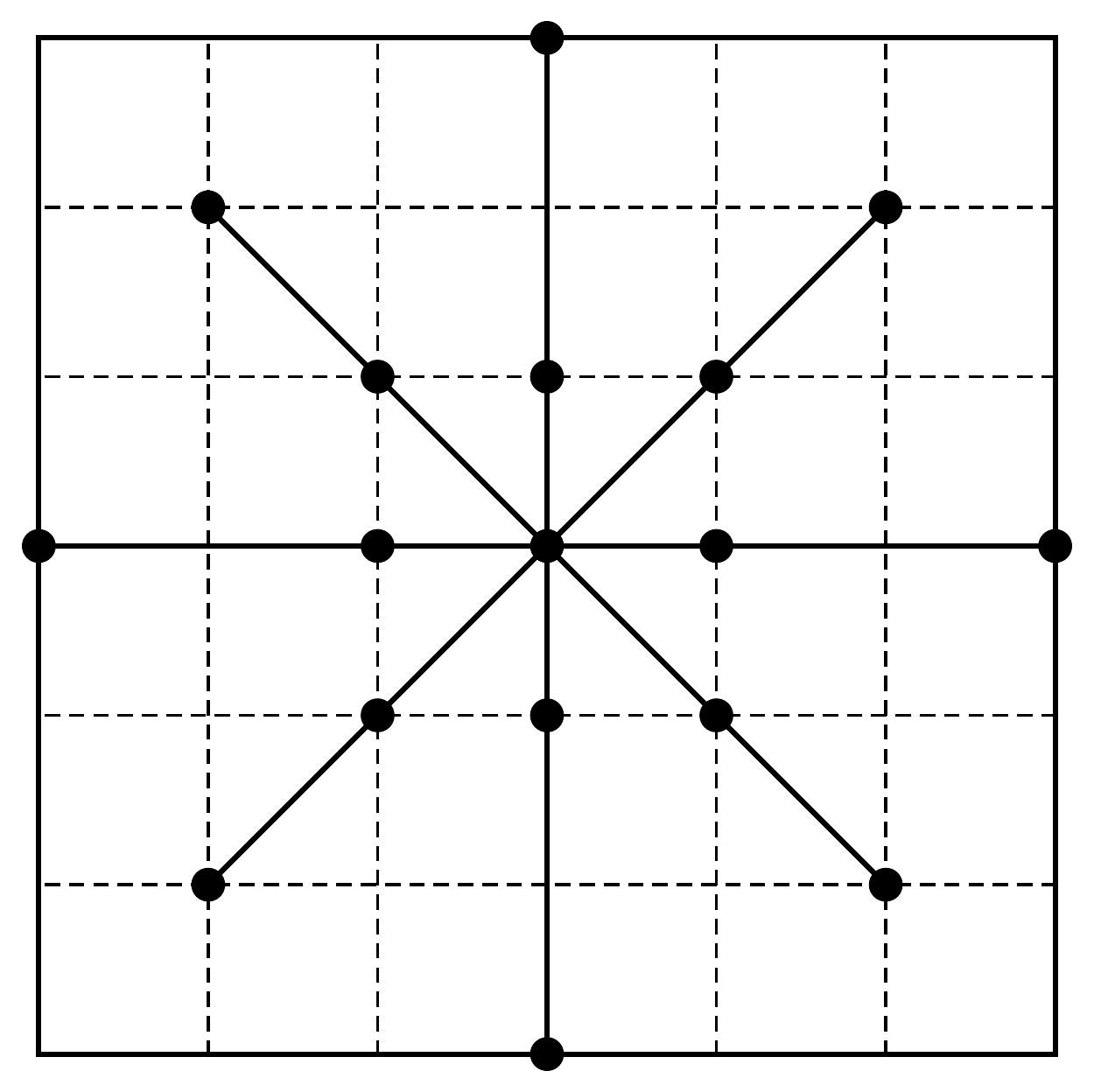}
\par\end{centering}

\caption{Illustration of the D2Q17 lattice. Each discrete velocity is represented
by the length and direction of the line connecting 
the origin of coordinate and the corresponding dotted point. \label{fig:D2Q17-Lattice}}

\end{figure}

\begin{figure}
\begin{centering}
\includegraphics[width=0.7\textwidth]{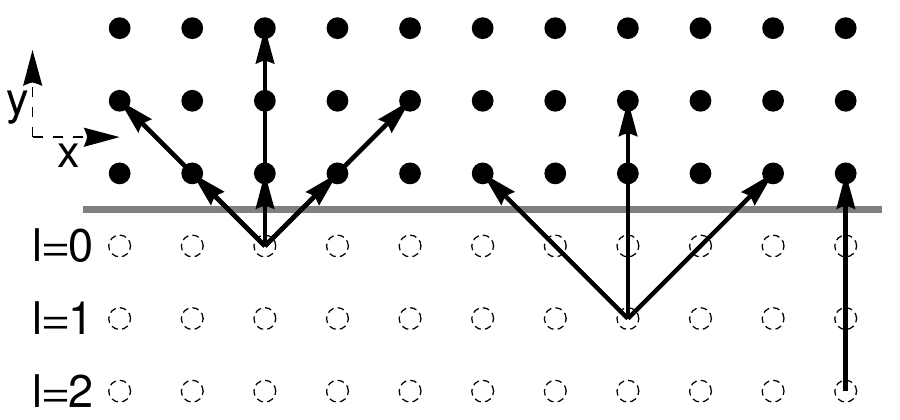}
\par\end{centering}

\caption{Illustration of the arrangement of the ghost grid and the corresponding
outgoing discrete velocity (arrow at bottom for the D2Q17 model (see
Fig.\ref{fig:D2Q17-Lattice}). The point denotes the bulk grid while
the dashed circle represents the ghost grid. \label{fig:Illustration}}
\end{figure}

To validate the proposed implementation of kinetic boundary condition,
we consider steady Couette flow confined in two parallel planar plates
located at $Y=0$ and $Y=1$ and moving oppositely with the same speed.
All the quantities are presented in their non-dimensional form, and
both isothermal and thermal conditions are considered. Therefore,
three lattice systems, namely D2Q17\cite{2006JFM550413S}, D2Q16\cite{Chikatamarla2006}
and D3Q121\cite{PhysRevE.81.036702,Nie2008} are tested, where the
D2Q17 and D2Q16 models are appropriate for the isothermal cases and
the D3Q121 model for the thermal ones. The D2Q17 lattice is illustrated
in Fig.\ref{fig:D2Q17-Lattice} and the corresponding grid arrangement
is shown in Fig.\ref{fig:Illustration}. The details of three lattices
are omitted here for simplicity, which can be found in \cite{2006JFM550413S},
\cite{Chikatamarla2006} and \cite{Nie2008}.

We restrict to the Couette flow within the slip-flow regime, so we
may be able to use solutions of the Navier-Stokes-Fourier (NS) equations as reference.
For the NS solutions, it is necessary to apply the velocity-slip
and temperature-jump boundary conditions so that the velocity and
temperature profiles can be written as

\begin{equation}
U_{NS}=\frac{(2y-1)}{2Kn+1}U_{w},
\end{equation}
and 
\begin{equation}
T_{NS}=\frac{Kn\left(8C_{p}T_{w}+5u_{w}^{2}\right)+8C_{p}T_{w}Kn^{2}+2C_{p}T_{w}-4PrU_{w}(Y-1)Y}{2C_{p}\left(2Kn+1\right)^{2}},
\end{equation}
where the Knudsen number is defined as 
\begin{equation}
Kn=\sqrt{\frac{\pi}{2}}\frac{\mu_{0}\sqrt{RT_{0}}}{p_{0}L}.
\end{equation}
$U_{w}$ denotes magnitude of the component of interest of the wall
velocity $\bm{u}_{w}$. For some relatively larger Knudsen numbers
we may also compare to the solution of the linearized Boltzmann-BGK
(L-BGK) equation. 

\begin{figure}
\begin{centering}
\includegraphics[width=0.47\textwidth]{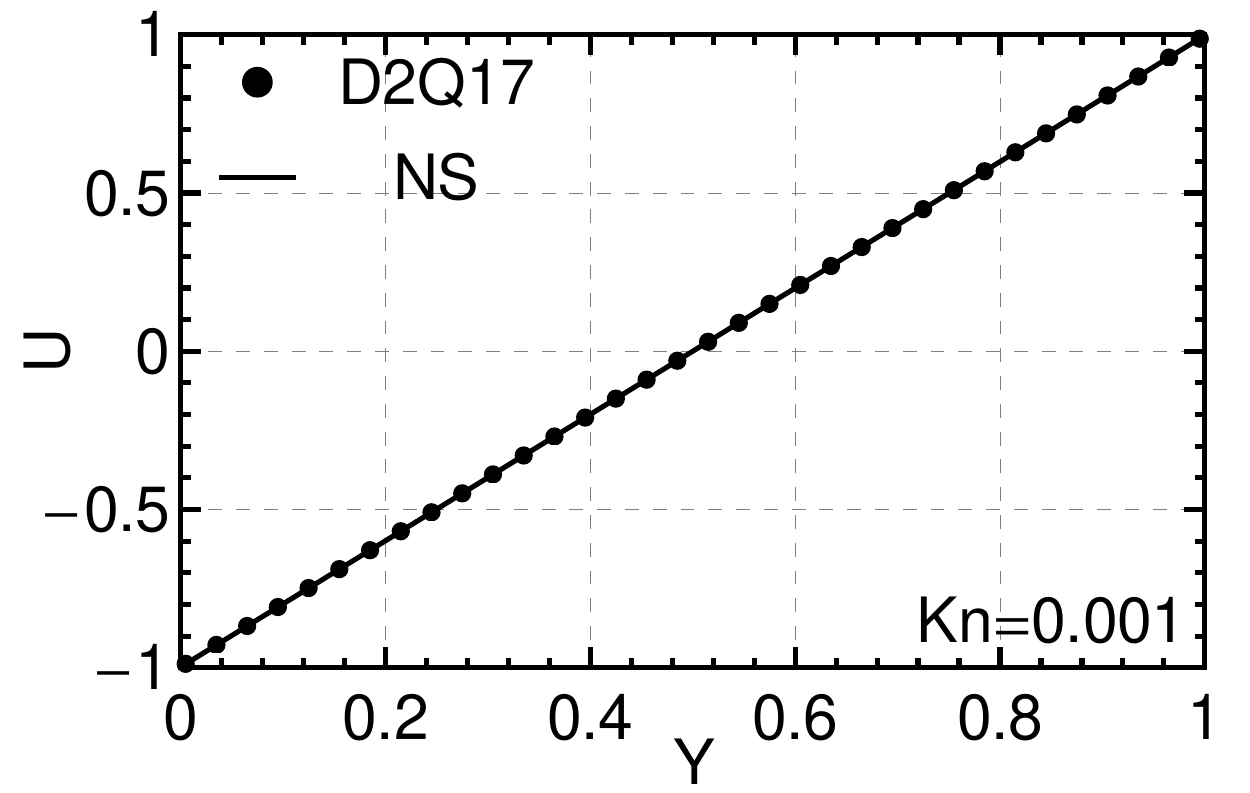}\includegraphics[width=0.47\textwidth]{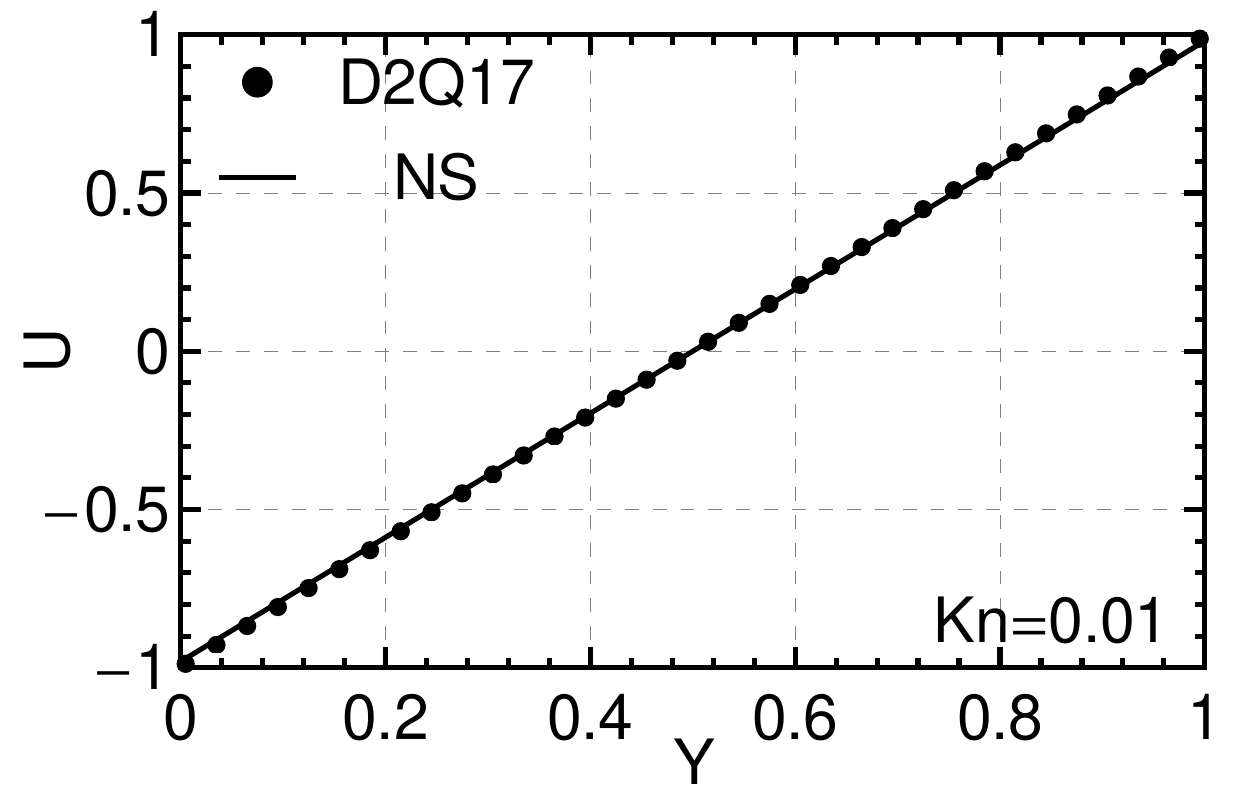}
\par\end{centering}

\begin{centering}
\includegraphics[width=0.47\textwidth]{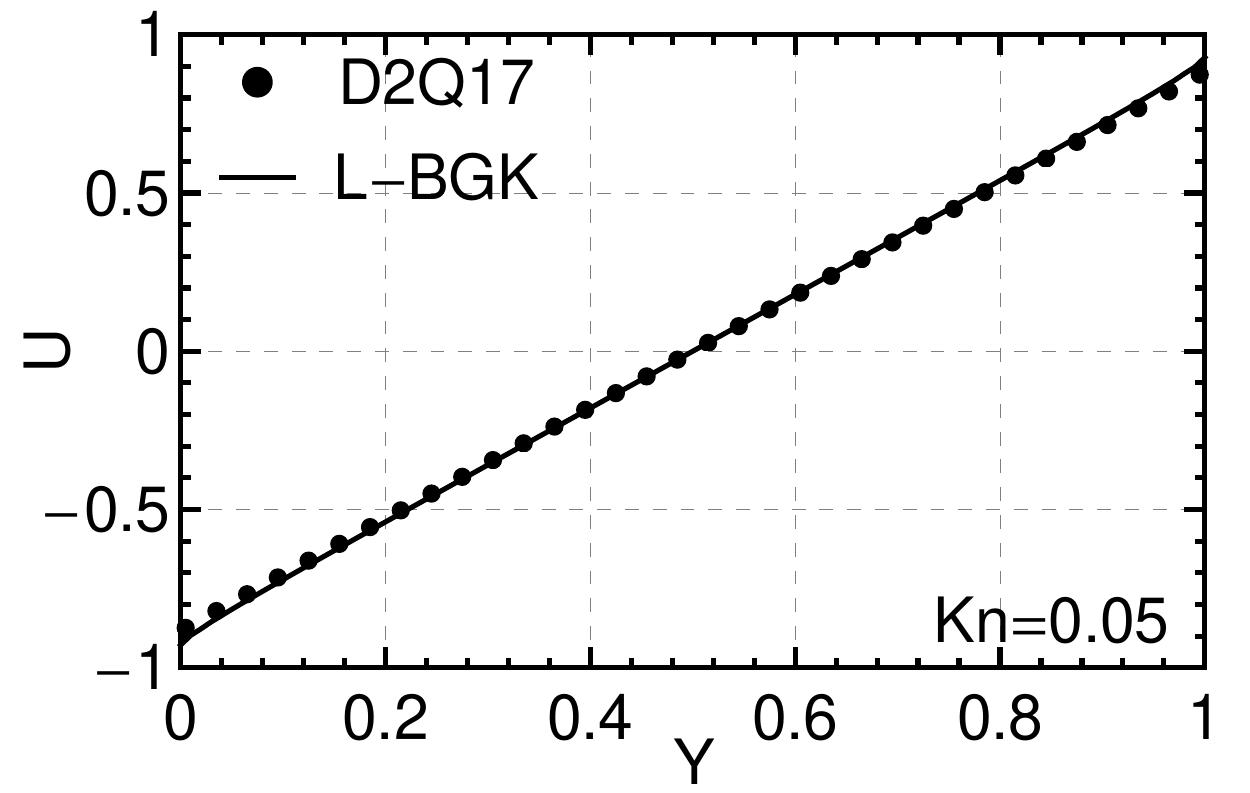}\includegraphics[width=0.47\textwidth]{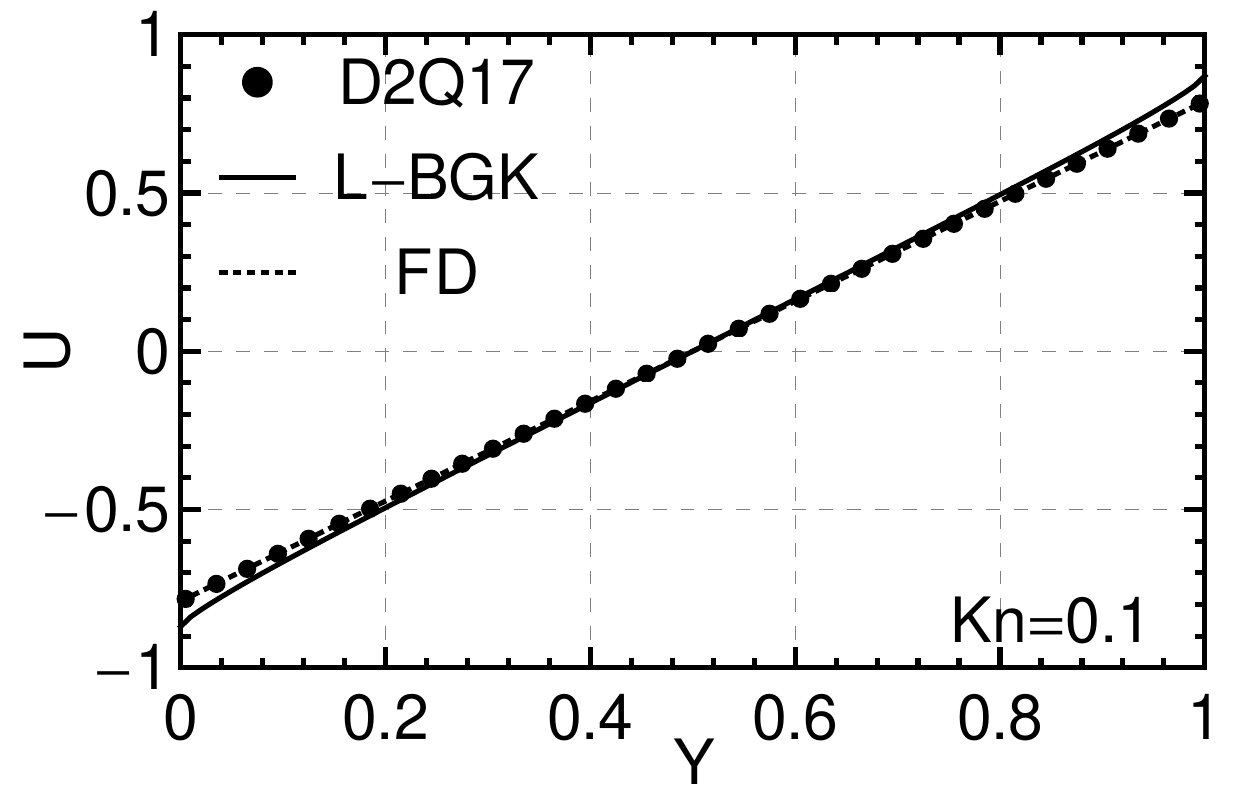}
\par\end{centering}

\caption{The velocity profiles for the isothermal cases with the D2Q17 model. The velocity
is further normalized by the wall velocity. \label{fig:isothermal-case}}
\end{figure}

\begin{figure}
\begin{centering}
\includegraphics[width=0.47\textwidth]{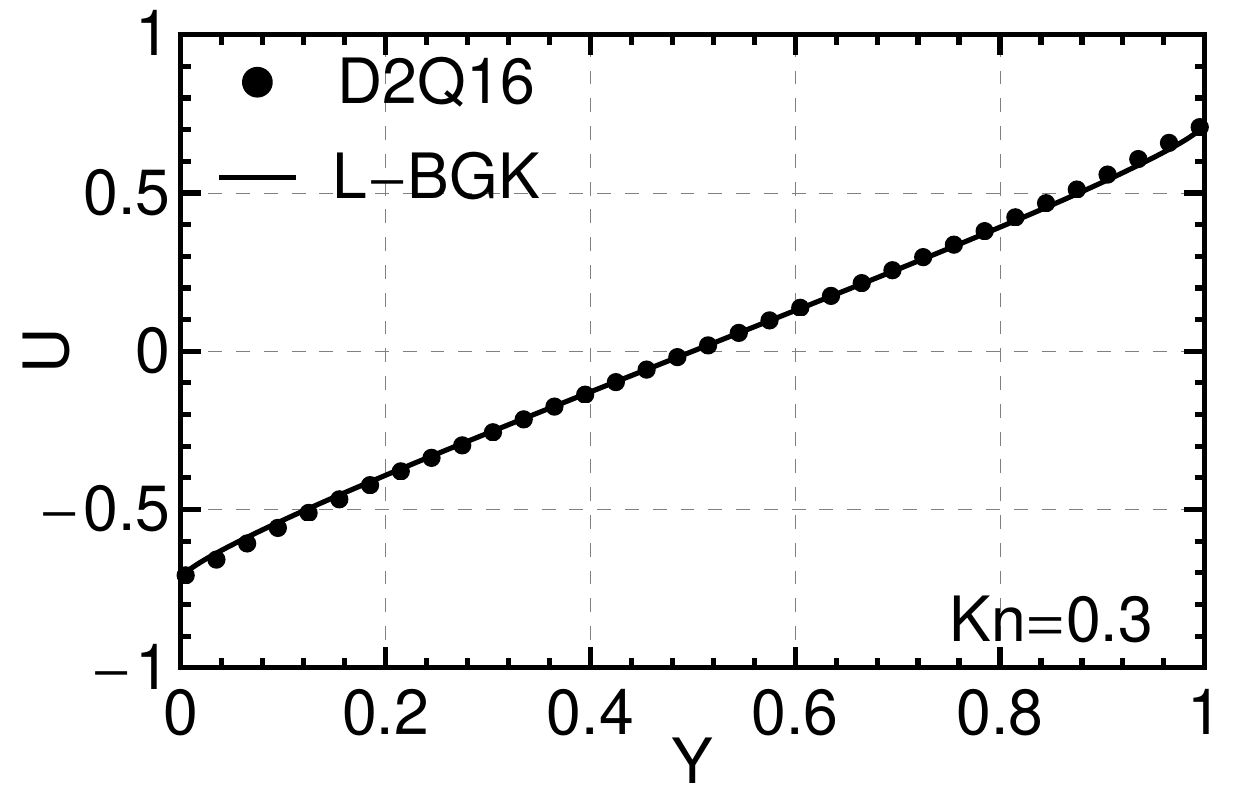}\includegraphics[width=0.47\textwidth]{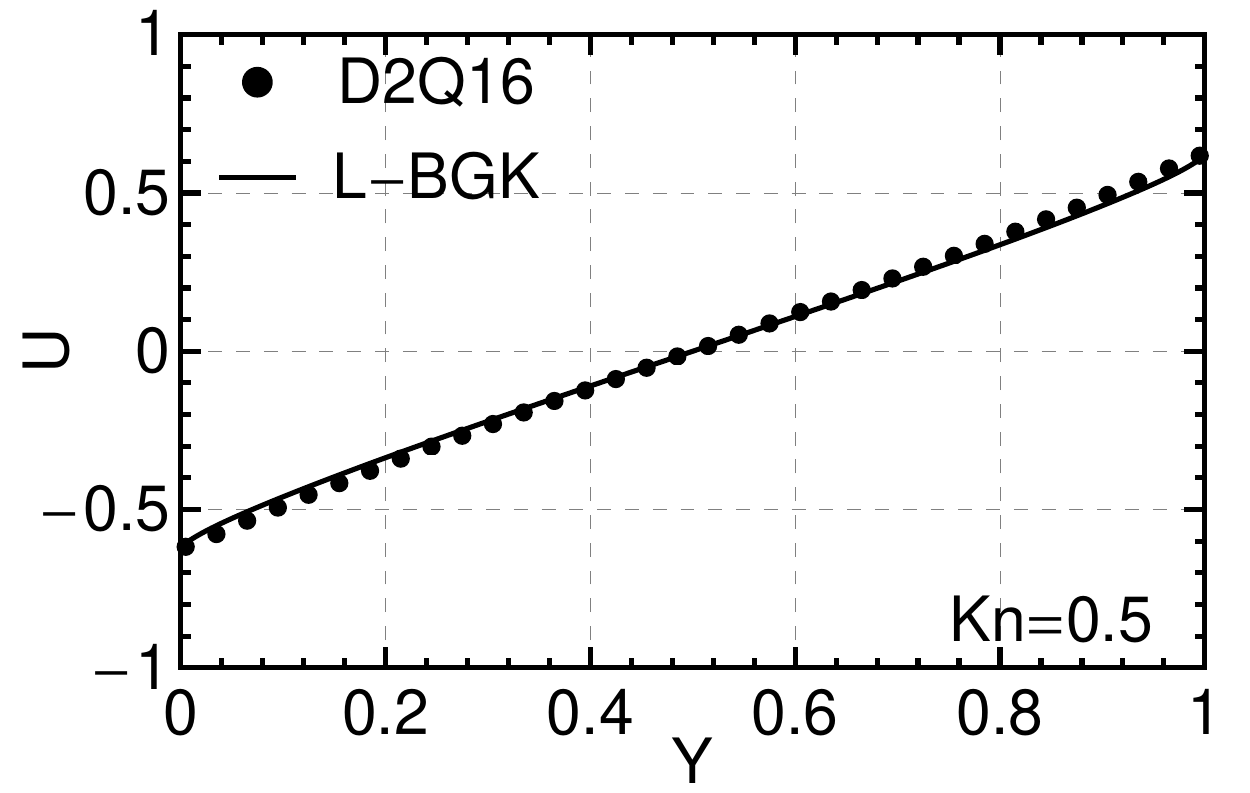}
\par\end{centering}

\caption{The velocity profiles for the isothermal cases with the D2Q16 model. The velocity
is further normalized by the wall velocity. \label{fig:isothermal-case-1}}
\end{figure}

We first evaluate the D2Q17 and D2Q16 models for isothermal flows
which are presented in Fig.\ref{fig:isothermal-case} and \ref{fig:isothermal-case-1},
where $U_{w}$ is set to be $0.05$. Both models are simulated with
$100$ computational grids in the direction of interest and the comparisons
are made against the NS solutions for $Kn<0.05$ and the L-BGK solutions
for $Kn\geq0.05$ respectively. The results show that the boundary
condition Eq.(\ref{eq:dkboundary}) works correctly for the isothermal
flows. For $Kn\leq0.05$, the velocity profiles are captured well
by the D2Q17 model while some deviations from the L-BGK results are
observed for larger Knudsen numbers, particularly at $Kn=0.1$ (see
Fig.\ref{fig:isothermal-case}). However, this is of no surprise as
it is known that these deviations are due to the lattice structure\cite{Meng2011b}.
A further comparison to the finite difference (FD) implementation
of Eq.(\ref{lbgk}) (see the description in \cite{Meng2011b}, where
the numerical simulation is validated by the analytical solution in
\cite{Ansumali2007}), confirms the appropriateness of the boundary
implementation. Interestingly, the D2Q16 model can given much better
predictions for the velocity profile. Even at $Kn=0.5$, it still gives
satisfactory results, see Fig. \ref{fig:isothermal-case-1}. The reason was already discussed in 
\cite{Meng2011b}


\begin{figure}
\begin{centering}
\includegraphics[width=0.47\textwidth]{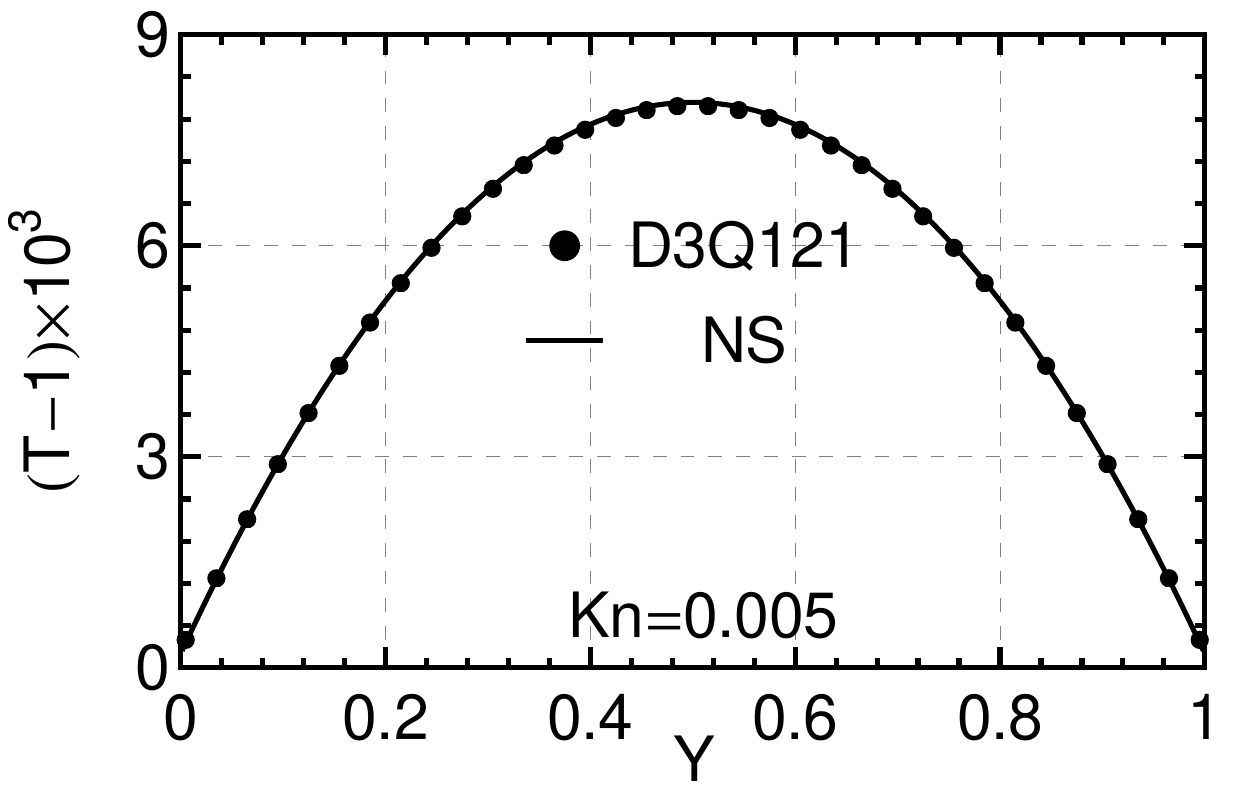}\includegraphics[width=0.47\textwidth]{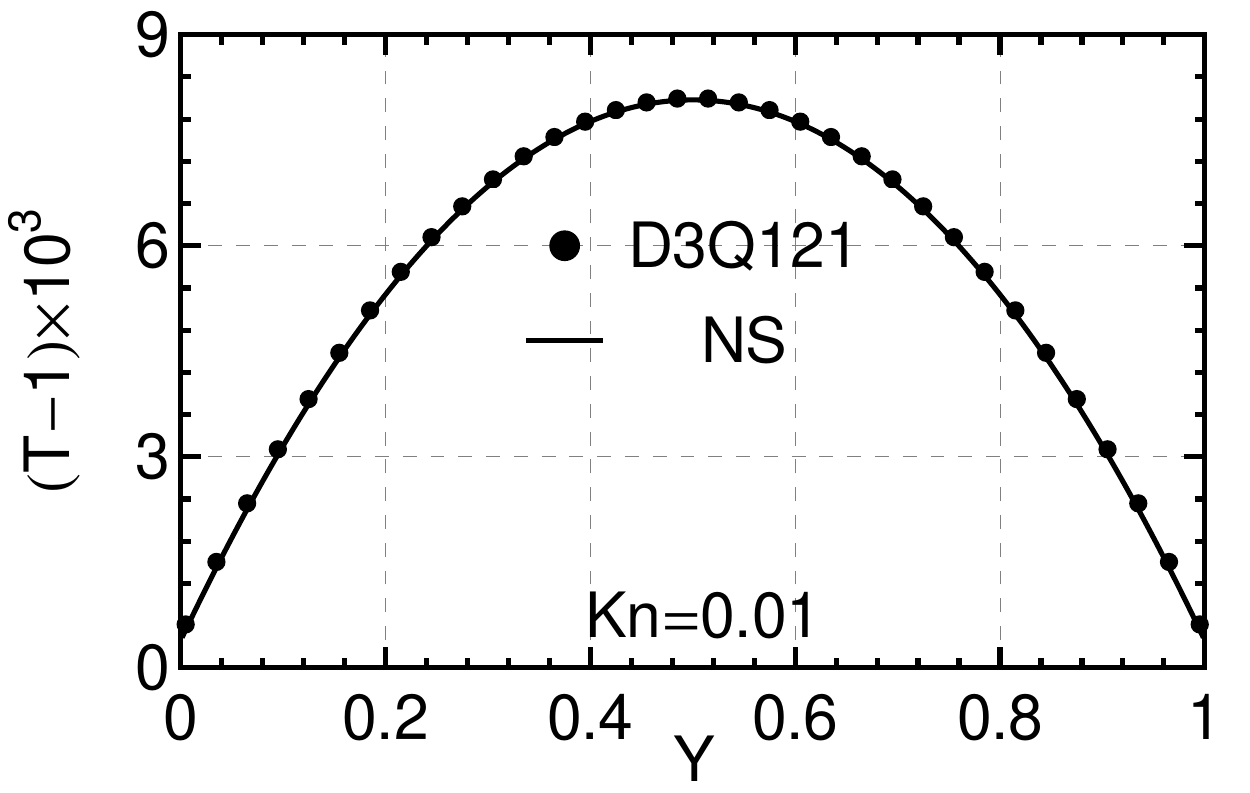}
\par\end{centering}

\caption{The temperature profiles for the thermal cases with $U_{w}=0.2$.\label{fig:thermal}}
\end{figure}

For further validation, we also simulate the thermal Couette flows
using the D3Q121 model. For these flows the relevant parameters are
$C_{p}=5/2$ and $Pr=1$ while the wall temperatures are set to be
$1$ and their speed $U_{w}$ is set to be $0.2$. As relatively small
Knudsen numbers are considered here, the results are compared to the
NS solutions, see Fig.\ref{fig:thermal}. The subtle temperature
jumps are well captured at the wall boundary. These agreements again
confirm the appropriateness of the boundary treatment. The velocity
profiles show similar behavior to the isothermal cases, so they are
not presented in Fig. \ref{fig:thermal}.

\begin{figure}
\begin{centering}
\includegraphics[width=0.47\textwidth]{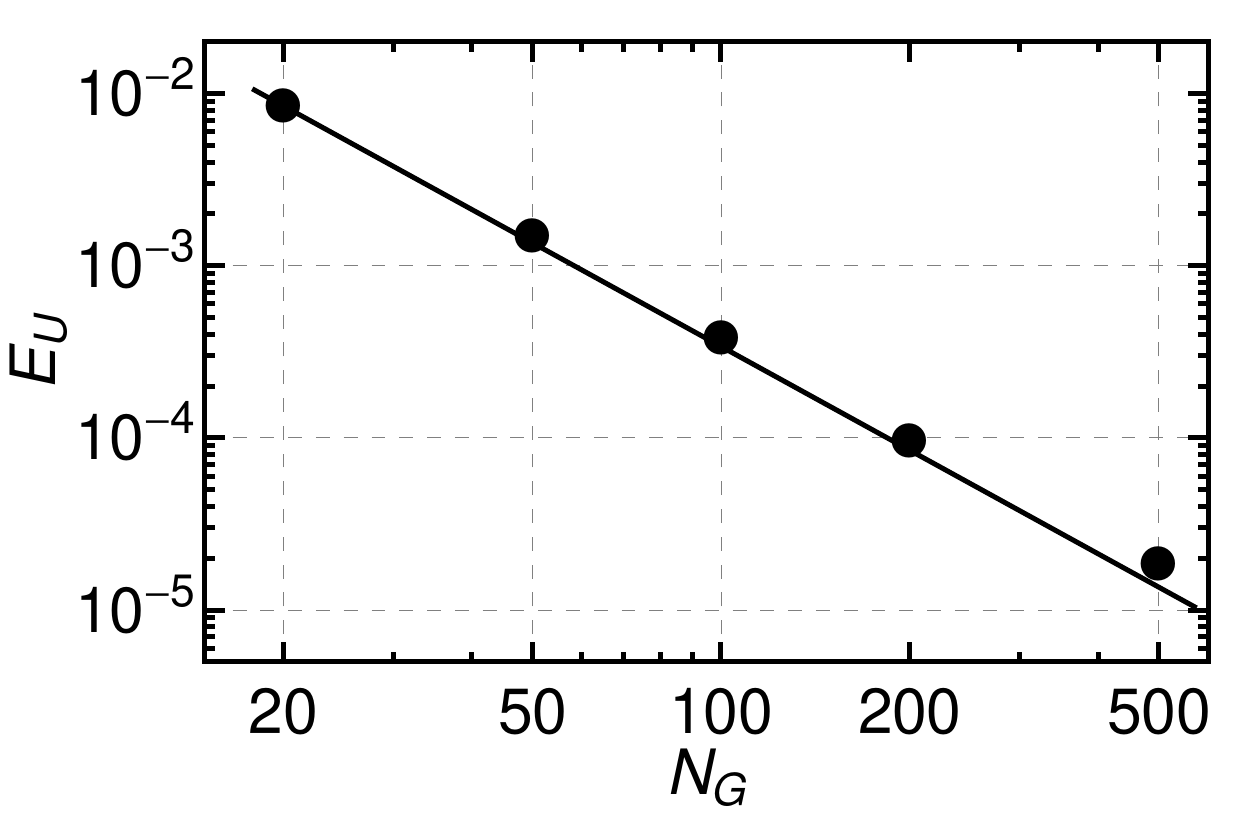}\includegraphics[width=0.47\textwidth]{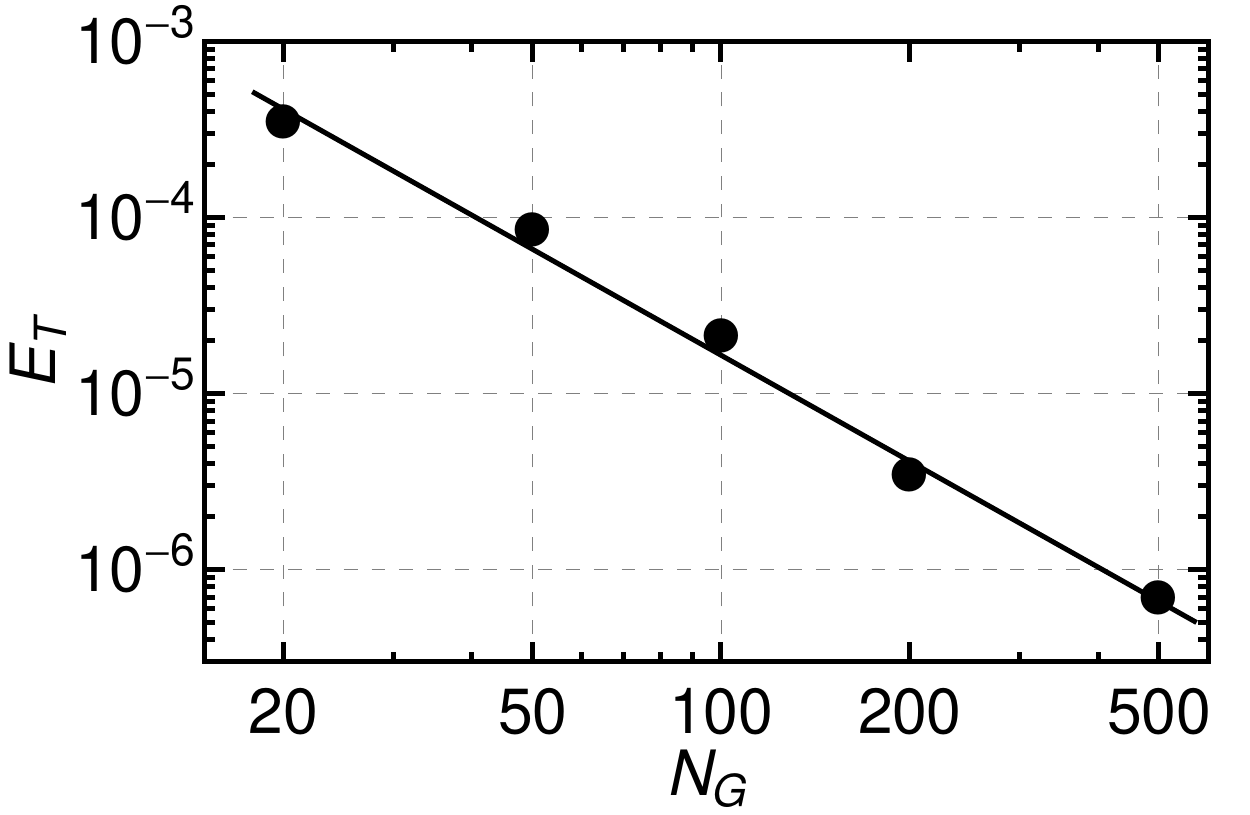}
\par\end{centering}

\caption{Dependence of the $E_{U}$ and $E_{T}$ on the grid number $N_{G}$.
\label{fig:logana}}
\end{figure}

To evaluate the numerical accuracy, a convergence study is conducted
for the thermal case of $Kn=0.01$. The simulations are run for six
different grid resolutions $N_{G}=20,50,100,200,500,1000$ in the direction of
interest. The results of $N_{G}=1000$ is then chosen as reference
and the global relative errors of the velocity and temperature are
defined as 
\begin{equation}
E_{U}=\sqrt{\frac{\sum_{j=1}^{N_{G}}[U_{j}(Y_{j})-U_{\Delta}(Y_{j})]^{2}}{\sum_{j=1}^{N_{G}}U_{\Delta}^{2}(Y_{j})}},
\end{equation}
and 

\begin{equation}
E_{T}=\sqrt{\frac{\sum_{j=1}^{N_{G}}[T_{j}(Y_{j})-T_{\Delta}(Y_{j})]^{2}}{\sum_{j=1}^{N_{G}}(T_{\Delta}(Y_{j})-T_{w})^{2}}},
\end{equation}
where $U_{\Delta}$ and $T_{\Delta}$ represent the results of $N_{G}=1000$.
Fig.\ref{fig:logana} shows that the second order accuracy is achieved
globally.

\section{Concluding remarks}

To conclude, we have formulated the kinetic diffuse reflection boundary
condition for high-order LB models with emphasis on retaining the
``streaming-collision'' mechanism. The numerical tests for
both isothermal and thermal Couette flows show that the present boundary
condition can capture velocity-slip and temperature-jump very well
within the capacity of the corresponding lattices. In term of numerical
accuracy, we show that the second order accuracy can be achieved globally.

\begin{acknowledgments}
The authors would like to thank Dr. Xiaowen Shan for many informative
discussions. The research leading to these results has received funding
from the Engineering and Physical Sciences Research
Council U.K. under Grants No. EP/F028865/1 and EP/ I036117/1.
\end{acknowledgments}

\end{document}